\documentclass{aa} % for a paper on 1 column  
\DeclareRobustCommand{\rchi}{{\mathpalette\irchi\relax}}
\newcommand{\irchi}[2]{\raisebox{\depth}{$#1\chi$}} % inner command, used by \rchi

\newcommand{\ex}{\mathbf{e}_{\rm x}}
\newcommand{\ey}{\mathbf{e}_{\rm y}}
\newcommand{\ez}{\mathbf{e}_{\rm z}}

\newcommand{\rlight}{r_{\rm L}}

\newcommand{\me}{m_{\rm e}}

\newcommand{\rot}{\mathbf{\nabla} \times}
\newcommand{\divg}{\mathbf{\nabla}\cdot}

\usepackage{graphicx}
\usepackage{txfonts}
\usepackage[colorlinks=true, linkcolor=blue, citecolor=blue, filecolor=blue, urlcolor=blue]{hyperref}

\begin{document} 

\title{Multi-wavelength emission in resistive pulsar magnetospheres}

%   \subtitle{I.}

\author{J. P\'etri
        %\inst{1}
         %\fnmsep\thanks{Just to show the usage of the elements in the author field}
       }

\institute{Universit\'e de Strasbourg, CNRS, Observatoire astronomique de Strasbourg, UMR 7550, F-67000 Strasbourg, France.\\
           \email{jerome.petri@astro.unistra.fr}         
}

\date{Received ; accepted }

% \abstract{}{}{}{}{} 
% 5 {} token are mandatory
 
\abstract
{Neutron star magnetospheres are well described in the two extreme cases of a vacuum field and a plasma filled force-free regime. However, none of these descriptions allow for magnetic field dissipation into particle kinetic energy and thus high-energy radiation. Some physical processes must be invoked to produce observational signatures typical of pulsars.}
{In this paper, we compute a full set of neutron star magnetosphere structures from the basic vacuum regime to the dissipation-less force-free regime by implementing a resistive prescription for the plasma. A comparison to the radiation reaction limit is also discussed. We investigated the impact of these resistive magnetospheres onto the multi-wavelength emission properties based on the polar cap model for radio wavelengths, on the slot gap model for X-rays and on the striped wind model for $\gamma$-rays.}
{We performed time-dependent pseudo-spectral simulations of the full Maxwell equations including a resistive Ohm's law. We deduced the polar cap shape and size, the Poynting flux, the magnetic field structure and the current sheet surface, depending on the magnetic obliquity~$\rchi$ and on the conductivity~$\sigma$.}
{We found that the geometry of the magnetosphere close to the stellar surface is not impacted by the amount of resistivity. Polar cap rims remain very similar in shape and size. However the Poynting flux varies significantly as well as the magnetic field sweep-back in the vicinity of the light-cylinder. This bending of field lines reflects into the $\gamma$-ray pulse profiles, changing the $\gamma$-ray peak separation~$\Delta$ as well as the time lag~$\delta$ between the radio pulse and $\gamma$-ray peaks. X-ray pulse profiles are also drastically affected by the resistivity.}
{A full set of multi-wavelength light-curves can be compiled for future comparison with the third $\gamma$-ray pulsar catalogue. This systematic study will help to constrain the amount of magnetic energy flowing into particle kinetic energy and shared by radiation.}


\maketitle

%
%-------------------------------------------------------------------

\section{Introduction}

Pulsars are detected in a broad band of the electromagnetic spectrum, from the radio wavelength up to very high energy photons in the GeV/TeV range. More than 300 $\gamma$-ray pulsars are known today \citep{smith_third_2023} for a total of about 4.400 radio pulsars listed in the online ATNF (Australia Telescope National Facility) pulsar catalogue\footnote{\url{https://www.atnf.csiro.au/research/pulsar/psrcat}} \citep{manchester_australia_2005}. Because this emission is pulsed, it must be produced close to the neutron star surface, in its magnetosphere and within its wind. Photons are produced by accelerating charged particle to ultra-relativistic speeds in the neutron star ultra-strong electromagnetic field. However, the details of this acceleration and radiation processes are still ill understood. The location of the emission sites are also loosely constrained. A quantitative understanding of the pulsar machinery requires a self-consistent modelling of the magnetosphere coupled to its emission mechanisms. Radiation implies that the star cannot be simply surrounded by empty space as in the Deutsch vacuum field solution \citep{deutsch_electromagnetic_1955}. In the other limit, it cannot contain an ideal plasma with infinite conductivity, pressure-less and with zero inertia leading to the force free regime because in such a regime, no dissipation is allowed. The truth lies between these two regimes, the magnetosphere must contain a plasma in a non-ideal regime, converting the stellar rotational kinetic energy into high frequency electromagnetic radiation in addition to the low frequency large amplitude electromagnetic wave launched by the rotating dipole.

Whereas the neutron star global electrodynamics is nowadays well reproduced thanks to numerical force-free simulations using finite difference schemes \citep{spitkovsky_time-dependent_2006} or pseudo-spectral algorithms \citep{petri_pulsar_2012, cao_spectral_2016}, the outcome in terms of realistic pulsed emission remained fuzzy. This leads several authors to consider resistive or dissipative magnetospheres in order to localize the sites where the energy transfer to particle and radiation occurs \citep{li_resistive_2012, palenzuela_modelling_2013, cao_oblique_2016, kalapotharakos_gamma-ray_2014, kato_global_2017, cao_three-dimensional_2020, petri_radiative_2020, petri_radiative_2022}. The most detailed picture is however given by kinetic simulations like particle in cell (PIC) codes, reaching close to force-free conditions as in \cite{philippov_ab_2015, cerutti_particle_2015} or reaching close to vacuum conditions as in \cite{mottez_ab_2024}. The main focus of most of these works \citep{bai_modeling_2010, kalapotharakos_gamma-ray_2012, brambilla_testing_2015, cerutti_modelling_2016, kalapotharakos_fermi_2017, cao_modeling_2019, yang_exploring_2021, cao_pulsar_2022, barnard_probing_2022, kalapotharakos_gamma-ray_2023, cao_modeling_2024} was on high-energy emission because at least for $\gamma$-ray pulsars, a large fraction of the spin-down luminosity flows into $\gamma$-ray photons. However, a multi-wavelength modelling effort, combining radio, X-ray and $\gamma$-ray knowledges emerged \citep{petri_multi-wavelength_2024-3, yang_modeling_2024} as a powerful tool to constrain even more the emission sites. Polarization is another important measurable quantity to constrain the geometry of the radiative sources within the magnetosphere \citep{cerutti_polarized_2016, petri_young_2021}.

Particle acceleration and very high-energy radiation production depend drastically on the plasma regime within the magnetosphere because the magnetic field aligned electric field strongly couples to the plasma screening efficiency. However, the exact plasma regime is still matter of debate especially as it can vary from place to place. The pair creation via photon disintegration in a strong magnetic field or via photon-photon interaction largely controls this regime as it was shown in PIC simulations \citep{chen_electrodynamics_2014, chen_filling_2020}.
The observational outcome of these various regimes should lead to different unambiguous imprints detectable in their multi-wavelength light-curves.

The aim of this paper is in a first step to compute force-free, radiative and resistive pulsar magnetospheres in the same framework, using the same pseudo-spectral numerical algorithm to compare the similarities and discrepancies between these assumptions. In a second step, we extract the observational signature by computing the multi-wavelength light-curves from radio to $\gamma$-rays, assuming a polar cap model for radio, a slot gap model for non-thermal X-rays and a striped wind model for the $\gamma$-rays. 
We stress that our emission models are solely based on geometrical considerations related to the magnetic field structure, ignoring the physical mechanism behind the radiation processes. Sec.\ref{sec:Model} presents the different regimes used to construct pulsar magnetospheres. Sec.\ref{sec:Magnetospheres} discusses the magnetospheric solutions found in the different regimes. The derived multi-wavelength emission properties are summarised in Sec.\ref{sec:Emission}. Conclusions are drawn in Sec.\ref{sec:Conclusion}.

\section{Magnetospheric plasma model\label{sec:Model}}

The vacuum and force-free dipole magnetospheres represent the two extreme models of a simple neutron star environment. The real magnetosphere lies somewhere in between those two limits, allowing for magnetic reconnection and dissipation. These non-ideal effects are described in its simplest form by a conductivity summarised in a single scalar parameter denoted by~$\sigma$. In this section, after a brief review on neutron star magnetospheric structure and emission sites, we remind the electric current prescription for the force-free, radiative and resistive plasma regimes, the vacuum current vanishing identically.

\subsection{Review on magnetospheric structure and emission}

Let us first briefly remind the qualitative structure of the magnetosphere and its related emission sites, see \cite{petri_theory_2016,cerutti_electrodynamics_2017, philippov_pulsar_2022} for recent reviews on their structure and dynamics and \cite{harding_pulsar_2017} for their emission models. The neutron star surface magnetic field is assumed dipolar, dragged by the stellar rotation, leading to regions of open and closed field lines. In the closed region, within the force-free regime, inside the light-cylinder of radius $\rlight$, the plasma corotates with the star and field lines remain inside $\rlight$. The open region corresponds to magnetic field lines crossing the light-cylinder and reaching very large distances. The surface separating the closed region from the open region is called the separatrix \citep{kalapotharakos_toward_2012}. Inside the light-cylinder, the north part of this separatrix joins the south part at the so-called Y-point exactly at the light-cylinder. The structure of this Y-point is still debated as it could in fact be a T-point, see for instance \cite{contopoulos_pulsar_2024} for an illustration.

\begin{figure}
	\centering
	\includegraphics[width=0.95\linewidth]{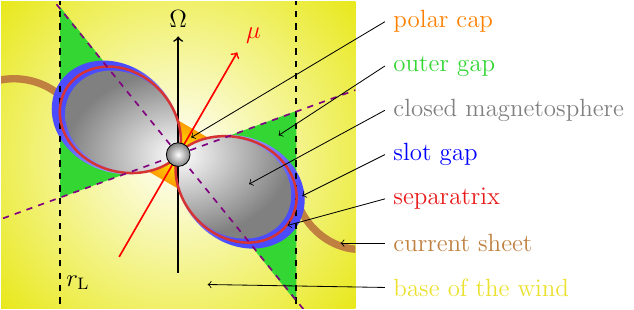}
	\caption{Schematic view of the pulsar magnetosphere, showing the separatrix and the emission sites.}
	\label{fig:magnetosphere}
\end{figure}

Based on this magnetospheric structure, three distinct regions were identified as potential emission sites for radio photons, X-rays and $\gamma$-rays. First the polar cap expected to produce the radio pulses, next the slot gap for non-thermal X-rays \citep{petri_localizing_2024} and the current sheet within the striped wind for $\gamma$-rays \citep{petri_unified_2011}. \cite{petri_multi-wavelength_2024-3} details these three emission models, for which even general-relativistic effects have been included by \cite{petri_general-relativistic_2018}. Figure~\ref{fig:magnetosphere} sketches the emission sites and magnetosphere topology for a general oblique rotator.

\subsection{Maxwell equations}

In all subsequent models, the plasma inertia and pressure is neglected. The plasma only furnishes the required charge~$\rho_{\rm e}$ and current~$\mathbf{j}$ densities to evolve Maxwell equations written in standard SI (Système International) units as
\begin{subequations}
	\begin{align}
		\label{eq:Maxwell1}
		\divg \mathbf B & = 0 \\
		\label{eq:Maxwell2}
		\rot \mathbf E & = - \frac{\partial \mathbf B}{\partial t} \\
		\label{eq:Maxwell3}
		\divg \mathbf E & = \frac{\rho_{\rm e}}{\varepsilon_0} \\
		\label{eq:Maxwell4}
		\rot \mathbf B & = \mu_0 \, \mathbf j + \frac{1}{c^2} \, \frac{\partial \mathbf E}{\partial t} \ .
	\end{align}
\end{subequations}
$\varepsilon_0$ is the vacuum permittivity, $\mu_0$ the vacuum permeability and $c$ the speed of light. Apart from the boundary conditions on the stellar surface and at infinity where outgoing wave conditions are imposed, the current density~$\mathbf{j}$ represents the only unknown of the problem. Once fixed according to a given plasma model, Maxwell equations are solved numerically, leading to a solution for the magnetosphere. So let us describe the simplest possibilities for this electric current density~$\mathbf{j}$.

\subsection{Current prescription}

We consider three prescriptions requiring either no free parameter as in the force-free approximation, or only one parameter, the pair multiplicity~$\kappa$ for the radiative model or a kind of conductivity parameter~$\sigma$ for the resistive model.

\subsubsection{Force-free regime}

The vacuum case is trivial with $\mathbf{j} = \mathbf{0}$ leading to the exact analytical solution given by \cite{deutsch_electromagnetic_1955}. In the exact opposite limit, an ideal plasma with infinite conductivity according to $\sigma=+\infty$, leads to the force-free prescription where the current density becomes \citep{gruzinov_stability_1999, blandford_lighthouse_2002} 
\begin{equation}
	\label{eq:J_ideal}
	\mathbf j = \rho_{\rm e} \, \frac{\mathbf{E}\wedge \mathbf{B}}{B^2} + \frac{\mathbf{B} \cdot \rot \mathbf{B} / \mu_0 - \varepsilon_0 \, \mathbf{E} \cdot \rot \mathbf{E}}{B^2} \, \mathbf{B} .
\end{equation}
This current does not produce work on particles since $\mathbf{j} \cdot \mathbf{E} = 0$ because $\mathbf{E} \cdot \mathbf{B} = 0$ by construction. All the rotational kinetic energy goes into the Poynting flux of the low frequency large amplitude electromagnetic wave. For completeness and faithful comparisons with other plasma regimes shown in this paper, we compute some force-free magnetosphere solutions with exactly the same numerical set up.

\subsubsection{Radiative approximation}

Going to a weakly dissipative regime, we prescribe the electric current associated to the radiation reaction limit velocity field. The detailed derivation is given by \cite{petri_theory_2016}, and the associated radiative current density~$\mathbf{j}$ with minimal assumption is explained in \cite{petri_radiative_2020}. The final expression reduces to
\begin{equation}
	\label{eq:J_rad}
	\mathbf j = \rho_{\rm e} \, \frac{\mathbf E \wedge \mathbf B}{E_0^2/c^2 + B^2} + |\rho_{\rm e}| \, (1 + 2\,\kappa) \, \frac{E_0 \, \mathbf E/c + c \, B_0 \, \mathbf B}{E_0^2/c^2 + B^2}
\end{equation}
where $\kappa$ is the pair multiplicity factor that controls the dissipation. This current converts the Poynting flux energy into particle kinetic energy because $\mathbf j \cdot \vec{E} > 0$ and subsequently into multi-wavelength high-energy radiation, essentially in X-rays and $\gamma$-rays.
The electric and magnetic field strengths, respectively $E_0$ and $|B_0|$ are related to the electromagnetic invariants\footnote{We impose $E_0>0$ and $B_0 = \textrm{sign}(\vec{E} \cdot \vec{B})|B_0|$.} by $E_0 \, B_0 = \vec{E} \cdot \vec{B}$ and $E_0^2 - c^2 \, B_0 = E^2 - c^2 \, B^2$. Physically, $E_0$ and $|B_0|$ represent the strength of the electric and magnetic field in any frame in which they are parallel.
However, in a series of previous works \citep{petri_electrodynamics_2020, petri_radiative_2020, petri_radiative_2022}
we found that the dissipation remains weak except along the separatrix and around the Y-point. The low dissipation rate arises because the plasma is always at least filled with the charge density $\rho_{\rm e}$ imposed by Maxwell-Gauss law Eq.~\eqref{eq:Maxwell3} combined to a non-negligible parallel current given by the second term in Eq.~\eqref{eq:J_rad} that never vanishes because $|\rho_{\rm e}| \, (1 + 2\,\kappa) \geq |\rho_{\rm e}|$. As already shown in these above-mentioned works \citep{petri_electrodynamics_2020, petri_radiative_2020, petri_radiative_2022}, this approximation is almost undistinguishable from the above force-free regime.

\subsubsection{Resistive model}

In order to circumvent this flaw of the radiative current, we tried another prescription allowing for a vanishing parallel current if $\sigma\to0$. The derivation of this current is given by \cite{li_resistive_2012} and reads
\begin{equation}
	\label{eq:J_res}
	\mathbf j = \frac{\rho_{\rm e} \, \mathbf{E}\wedge \mathbf{B} + \Gamma \, \sigma \, E_0 \, ( E_0 \, \mathbf E/c^2 + B_0 \, \mathbf B)}{E_0^2/c^2 + B^2}
\end{equation}
with \begin{equation}\label{key}
	\Gamma = \sqrt{\frac{c^2\,B^2+E_0^2}{c^2\,B_0^2+E_0^2}} \ .
\end{equation}
The previous factor $|\rho_{\rm e}| \, (1 + 2\,\kappa)$ in Eq.~\eqref{eq:J_rad} is now replaced by $\Gamma \, \sigma \, E_0 / c$ in Eq.~\eqref{eq:J_res}. As will be shown in the next section, the limit $\sigma\to0$ corresponds to the vacuum solution whereas the $\sigma\to+\infty$ limit corresponds to the force-free regime. Therefore, $\sigma$ can be interpreted as a kind of conductivity controlling the rate of magnetic dissipation into plasma heating and acceleration that is however not catch in these models. We emphasise that eq.~\eqref{eq:J_res} is a special case, called minimal velocity limit, of a more general expression derived by \cite{li_resistive_2012} which includes a velocity component of the fluid $v_\parallel$ along the common direction of $\vec{E}_0$ and $\vec{B}_0$. However, imposing $v_\parallel \neq 0$ does not guaranty to tend to the vacuum case whenever $\sigma\to0$ and therefore $\sigma$ can no longer be interpreted as a conductivity if $v_\parallel \neq 0$. If the limiting case $v_\parallel=\pm c$ is applied in the general prescription of the current density in \cite{li_resistive_2012}, it would reduce to some formal expression very similar to Eq.\eqref{eq:J_rad} with no dependence on the conductivity $\sigma$. For instance imposing $v_\parallel = -\textrm{sign}(\rho_{\rm e}) \, c$, corresponding to instantaneous acceleration to the speed of light in the direction of the comoving electric field (see Eq.(30) in \cite{petri_electrodynamics_2020}), leads exactly to Eq.\eqref{eq:J_rad} with $\kappa=0$.

\section{Magnetospheric solutions\label{sec:Magnetospheres}} %Energetics}

In this section we present the numerical results of resistive and radiative magnetosphere solutions obtained by our pseudo-spectral time evolution code using spherical polar coordinated $(r,\theta,\varphi)$, see \cite{petri_pulsar_2012} for details of implementation and \cite{petri_general-relativistic_2014} for its extension to general-relativity.
In order to get more insight into these results, we first show the spin down luminosity depending on the obliquity and plasma regime, then the magnetic field structure and finally the polar cap shapes and the separatrix. % region of strong magnetic dissipation. 
This will serve as the basis for the discussion about the multi-wavelength emission properties of resistive magnetospheres.
For numerical purposes, we set the inner boundary of the simulation domain to $R_1/\rlight=0.2$, the outer boundary to  $R_2/\rlight=8$ where $\rlight = c/ \Omega$ is the light-cylinder radius and $\Omega$ the neutron star angular velocity. The numerical grid size is $N_r \times N_\theta \times N_\varphi = 129\times32\times64$. The simulations run on a single CPU machine and takes on average one week. $R_1$ is identified as being the neutron star radius~$R$. The most dissipative radiative case is given by $\kappa=0$ and will be the only radiative model considered in this work. The case $\kappa=0$ means that no pairs are created and that the plasma is fully charge-separated, only primary particles contribute to the current and charge density. Dissipation of the electromagnetic wave is largest in this case. However this regime is still far from vacuum as particles populate the whole magnetosphere, mimicking an almost force-free regime.
For the conductivity~$\sigma$ we choose value such that it smoothly transitions between the vacuum and force-free case, taking $\log (\sigma/\varepsilon_0 \, \Omega) = \{-1,0,1,2,3\}$. The normalised conductivity parameter is defined by~$\tilde{\sigma} = \sigma/\varepsilon_0 \, \Omega$. The obliquity of the pulsar is chosen to be $\rchi=\{0\degr, 15 \degr, 30\degr, 45\degr, 60\degr, 75\degr, 90\degr\}$.

\subsection{Spin down luminosity}

The fundamental quantity controlling the secular evolution of the neutron star magnetosphere is its spin down, the rate of energy radiated by the electromagnetic wave. As a reference value, we compute the luminosity through the Poynting flux for a perpendicular point dipole rotating in vacuum and given by \cite{deutsch_electromagnetic_1955}
\begin{equation}\label{eq:Lperp}
	L_\perp^{\rm vac} = \frac{8\,\pi}{3\,\mu_0\,c^3}\,\Omega^4\,B^2\,R^6 ,
\end{equation}
where $B$ is the magnetic field strength at the equator of a spherical star, $\Omega$ is its rotation rate, and $R$ is its radius. Expression~\eqref{eq:Lperp} is valid for a point dipole or in the limit $R\ll\rlight$. If $R\lesssim\rlight$ then some corrections including spherical Hankel functions need to be applied, see \cite{petri_multipolar_2015}, particularly eq.(46).

\begin{figure}[h]
	\centering
	\includegraphics[width=0.95\linewidth]{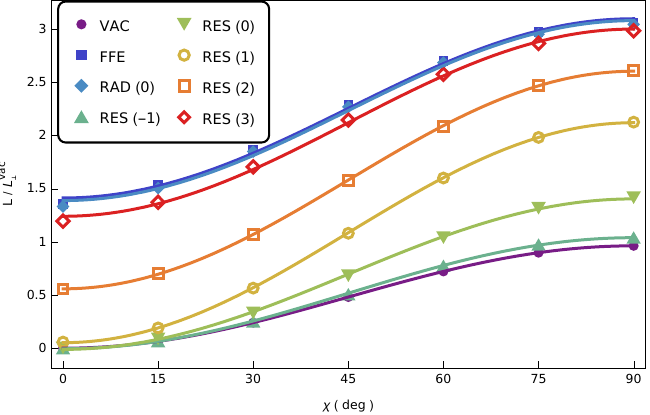}
	\caption{Spin down luminosity depending on the obliquity~$\rchi$ of the magnetosphere. The vacuum (VAC) and force-free (FFE) cases are shown as references.} 
	\label{fig:spindown}
\end{figure}

Fig.~\ref{fig:spindown} shows the spin down luminosity~$L$ depending on the magnetic obliquity~$\rchi$ and on the plasma regime: VAC for the vacuum, FFE for the force-free, RAD for the radiative regime with $\kappa=0$ and RES for the resistive plasma with the value of $\log \tilde{\sigma}$ in parentheses. As expected, for a given obliquity,  all luminosities vary between the vacuum and the force-free case, the vacuum being reached for $\log\tilde{\sigma}\ll1$ whereas the force-free regime corresponds to the limit $\log \tilde{\sigma}\gg1$. All solutions show a $\sin^2\rchi$ dependence with respect to the obliquity~$\rchi$. The results can be summarised by the expression
\begin{equation}\label{eq:spindown_fit}
	\frac{L}{L_\perp^{\rm vac}} = a + b \, \sin^2 \rchi
\end{equation}
with the value of the coefficients $a$ and $b$ given in Table~\ref{tab:coeff}. Note that in the vacuum case, $a=0$ but $b\lesssim1$ because of the large ratio $R/\rlight=0.2$, leading to a correction to leading order of this factor~$b$ by an amount $b\approx 1-(R/\rlight)^2 \approx 0.96$.
\begin{table}[h]
	\centering
	\caption{Analytical fits to the spin down luminosities, in the form of Eq.~\eqref{eq:spindown_fit} as $L/L_\perp^{\rm vac} = a + b \, \sin^2 \rchi$.\label{tab:coeff}}
	\begin{tabular}{lcc}
		\hline
		& a & b \\
		\hline
		VAC     & $0.00$ & $0.96$ \\
		RES (-1) & $0.00$ & $1.04$ \\
		RES (0)  & $0.00$ & $1.41$ \\
		RES (1)  & $0.05$ & $2.06$ \\
		RES (2)  & $0.56$ & $2.04$ \\
		RES (3)  & $1.24$ & $1.76$ \\
		RAD (0)  & $1.39$ & $1.69$ \\
		FFE     & $1.41$ & $1.69$ \\
		\hline
	\end{tabular}
\end{table}

\begin{figure}[h]
	\centering
	\includegraphics[width=0.95\linewidth]{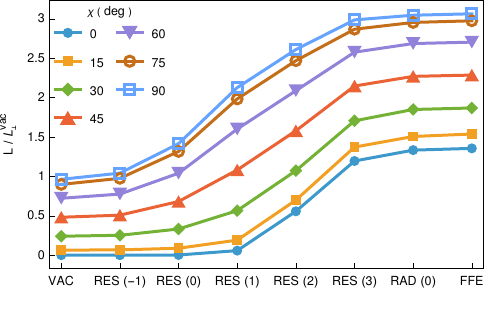}
	\caption{Spin down luminosity for fixed obliquity~$\rchi$ of the magnetosphere. The vacuum (VAC) and force-free (FFE) cases are shown as the limiting cases.}
	\label{fig:spindown2}
\end{figure}

Fig.~\ref{fig:spindown2} shows the same results but from a different perspective, changing the plasma regime and fixing the obliquity $\rchi$. The transition from VAC to FFE solution is clearly seen. The heuristic parameter $\sigma$ is helpful for controlling the plasma regime from non dissipation to maximal dissipation even if it hides all the complex microphysics of particle dynamics and radiation that can only be caught by performing detailed kinetic simulations.

\subsection{Particle versus Poynting luminosity}

\begin{figure}[h]
	\centering
	\includegraphics[width=0.95\linewidth]{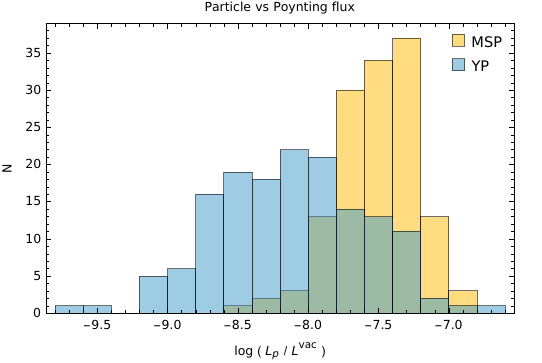}
	\caption{The ratio ${L_{\rm p}}/{L^{\rm vac}_\perp}$ for $\Gamma=10$ and $\kappa=0$ for millisecond pulsars (MSP) and young pulsars (YP) according to data from 3PC.}
	\label{fig:particule_vs_poynting_flux}
\end{figure}

The luminosity computed in our model takes only the Poynting flux into account. There is no contribution from the particle neither in the force-free limit nor in the resistive regime because particles possess zero inertia. They only contribute to charge and current densities. Therefore, strictly speaking they do not take away any energy from the neutron star. However, because of their mass, their number density and relativistic speed, there is actually a particle kinetic energy flux that amounts to approximately
\begin{equation}
	L_{\rm p} = 8\,\pi \, \varepsilon_0 \, \Gamma \, (1+2\,\kappa) \, \frac{\me \,c}{e} \, \Omega^2 \, B \,R^3 
\end{equation}
assuming a particle number density depending on the Goldreich-Julian charge separated density (of the order $2\,\varepsilon_0\,\Omega\,B$), a pair multiplicity factor~$\kappa$ and a bulk Lorentz factor~$\Gamma$ for the cold wind. We neglect the contribution from the hot current sheet, see for instance \cite{petri_illusion_2019}. This energy flux remains negligible as long as the ratio
\begin{equation}
	\frac{L_{\rm p}}{L^{\rm vac}_\perp} = 3 \, \Gamma \, (1+2\,\kappa) \, \frac{\me \,c^3}{e\,B\,\Omega^2\,R^3}
\end{equation}
remains much smaller than unity. This ratio is shown in Fig.~\ref{fig:particule_vs_poynting_flux} for $\Gamma=10$ (mildly relativistic flow) and $\kappa=0$ (no pair creation, fully charge-separated plasma current) for the $\gamma$-ray pulsars in the third catalogue 3PC \citep{smith_third_2023}, separately for millisecond pulsars (MSP) and young pulsars (YP). This ratio is always smaller than $10^{-7}$, thus even for typical values for the bulk flow Lorentz factor of $\Gamma\lesssim100$ and pair multiplicity of $\kappa\lesssim10^5$, this ratio is smaller than one. The energy taken away by the particle represents at most several percents of the Poynting flux.

\subsection{Magnetic field}

\begin{figure}[h]
	\centering
	\includegraphics[width=0.95\linewidth]{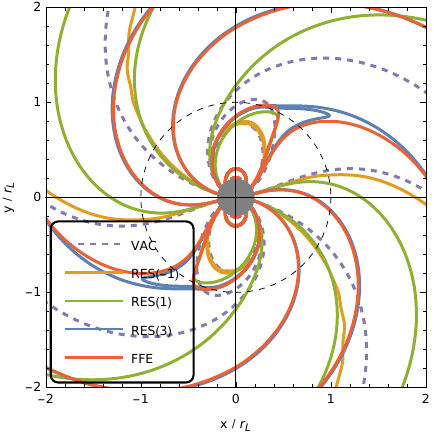}
	\caption{Magnetic field lines in the equatorial plane of an orthogonal rotator for different plasma regimes going from VAC to FFE through RES. The dashed circle depicts the light-cylinder radius and the gray disk the neutron star.}
	\label{fig:ligneschampbxyr0}
\end{figure}

The resistivity impacts significantly the geometry of the magnetic field at large distances from the neutron star surface. The vacuum and force-free field lines are very different at the light cylinder as seen in Fig.~\ref{fig:ligneschampbxyr0} for a perpendicular rotator. All other possible geometries between these two limiting cases are found depending on the conductivity~$\sigma$. The presence of a plasma within the magnetosphere forces the field lines to sweep backwards stronger than in the vacuum regime. For instance the field line rooting at the magnetic pole crosses the light-cylinder at later rotational phase when $\sigma$ increases to the point where it joins the force-free limit. As a consequence, the curvature of the field lines also increases because of the plasma back reaction. As the $\gamma$-ray emission is expected to be produced in the vicinity of the light-cylinder, this sweep back has profound impact on the radio to $\gamma$-ray time lag as observed in the third pulsar catalogue (3PC), \citep{smith_third_2023}. We discuss this in depth in section~\ref{sec:Emission}. The RAD results are very similar to the FFE results and therefore not shown for readability reasons.

\begin{figure}[h]
	\centering
	\includegraphics[width=0.95\linewidth]{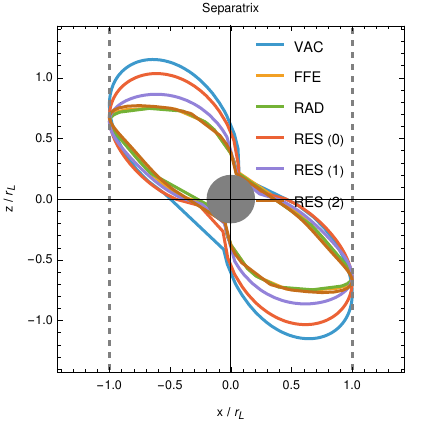}
	\caption{Separatrix cross-section in the $\vec{\mu}-\vec{\Omega}$ plane for an obliquity ${\rchi=45\degr}$ and different plasma regimes. The dashed vertical lines depict the light-cylinder and the gray disk the neutron star.}
	\label{fig:separatrice}
\end{figure}

The separatrix represents an important surface within the magnetosphere. It is significantly impacted by the plasma resistivity. 
Fig.~\ref{fig:separatrice} shows an example of cross-section in the $\vec{\mu}-\vec{\Omega}$ plane, i.e. the plane defined by the magnetic moment vector $\vec{\mu}$ and the rotation axis $\vec{\Omega}$, for an obliquity $\rchi=45\degr$ and for different values of the conductivity~$\sigma$. An increase in this conductivity shrinks the volume of the separatrix. We note that the location where this separatrix touches the light-cylinder does not significantly depend on the plasma regime. This is an important point since these places are the base of the current sheet in the striped wind. The low altitude region of the separatrix also delimits the radio beam opening angle and recent studies \citep{petri_localizing_2024} suggest that non-thermal X-ray emission is supported by this surface. Therefore, the separatrix is at the heart of the multi-wavelength pulsed emission of pulsars.

\subsection{Polar caps}

\begin{figure}[h]
	\centering
	\includegraphics[width=0.95\linewidth]{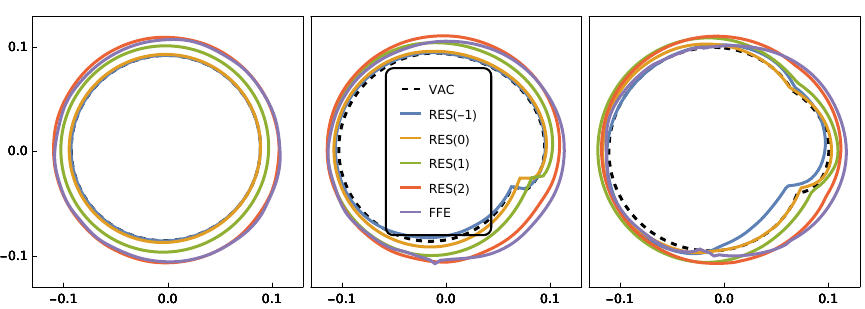}
	\caption{Polar cap shape viewed from above for $\rchi = \{15\degr, 45\degr, 75\degr\}$, from left to right. The vacuum case is shown in black dashed lines.}
	\label{fig:polar_cap_resistif_r0.2_a15}
\end{figure}

The structure of the wind and its current sheet rely on the shape of the separatrix rooted to the polar cap geometry. It is therefore important to localize on the neutron star surface the feet of the last closed field lines separating the inert corotating closed magnetosphere from the out-flowing plasma responsible for radio and high-energy emission. We therefore computed the polar cap rims in the resistive regime for any obliquity and conductivity. Results for $\rchi = \{15\degr, 45\degr, 75\degr\}$ are shown in Fig.~\ref{fig:polar_cap_resistif_r0.2_a15}. The vacuum rotator produces the smallest size polar caps whereas the force-free regime produces mainly the largest caps. For high inclinations, the vacuum case produces artificial cusps in the polar caps that disappear in a plasma loaded magnetosphere. In this regime, all the polar cap rims tend to circularise independently of the obliquity.

\begin{figure}[h]
	\centering
	\includegraphics[width=0.95\linewidth]{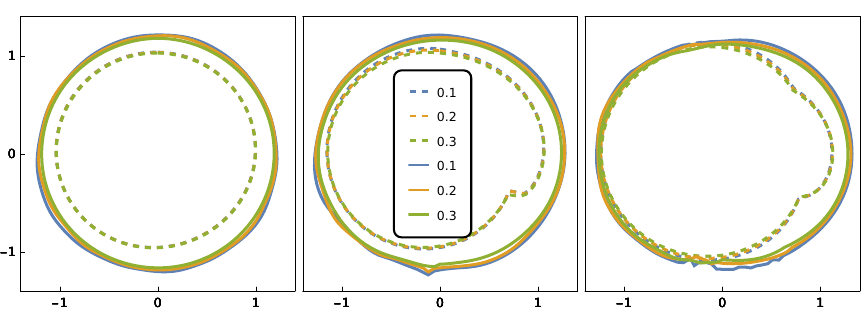}
	\caption{Polar cap shape for different obliquities $\rchi = \{15\degr, 45\degr, 75\degr\}$, from left to right, and several ratio $R/\rlight=\{0.1,0.2,0.3\}$. The sizes are normalised according to the scaling with respect to $\sqrt{R/\rlight}$, thus the horizontal $x$-axis is $(x/R)$ and the vertical $y$-axis is $(y/R)$ in units of  $\sqrt{R/\rlight}$.} 	\label{fig:polar_cap_rayon_r0.2_a15}
\end{figure}

The shapes found correspond to an artificially large neutron star radius with $R/\rlight=0.2$. This limitation is due to the computational resources needed to compute magnetosphere solutions with smaller stellar radii. However, we checked that the polar cap rims previously shown in Fig.~\ref{fig:polar_cap_resistif_r0.2_a15} scale with the theoretical ratio $\sqrt {R/\rlight}$ for $R/\rlight\ll1$. To verify this assertion, we plot the polar cap rims in the extreme limit of vacuum polar caps in dashed lines, resistive polar caps and force-free polar caps in solid lines in Fig.~\ref{fig:polar_cap_rayon_r0.2_a15} for different values of the ratio $R/\rlight=\{0.1,0.2,0.3\}$. For smaller ratios, the agreement would be even better as can be checked accurately in the vacuum case for any $R/\rlight$ (for which analytical solutions exists).

\begin{figure}[h]
	\centering
	\includegraphics[width=0.95\linewidth]{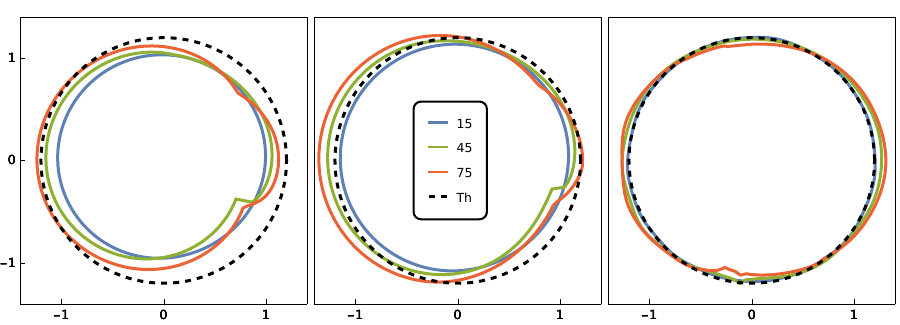}
	\caption{Polar cap rim centred onto the magnetic pole for different inclination angles~$\rchi$ in the vacuum, left panel, resistive, middle panel, and force-free regime, right panel. The fit Eq.~\eqref{eq:polar_cap_radius} is shown as a dashed black circle of normalised radius $1.2$.}
	\label{fig:polar_cap_r0.1_angle}
\end{figure}

Moreover, in the force-free regime, to very good accuracy, the polar cap rim is insensitive to the obliquity $\rchi$ as shown in Fig.~\ref{fig:polar_cap_r0.1_angle}. It is accurately approximated by a constant radius $r_{\rm pc}$ independent of $\rchi$ such that 
\begin{equation}\label{eq:polar_cap_radius}
	r_{\rm pc}/\rlight \approx 1.2 \, (R/\rlight)^{3/2} \qquad \textrm{or} \qquad r_{\rm pc}/R \approx 1.2 \, \sqrt{R/\rlight} \ .
\end{equation}
This relation can be used as a good proxy to estimate the size of polar caps deduced from the dipolar magnetic field.

\section{Multi-wavelength pulsed emission\label{sec:Emission}}

The above considerations described the magnetic environment of the neutron star and their magnetic topology, focusing on important concepts like spin down luminosity, polar caps and separatrix surface. However, such configurations are not easily uncovered by direct measurements. We need to rely on indirect inference by observing their pulsed electromagnetic emission. Therefore, in this section, we investigate in detail the multi-wavelength outcome of this magnetospheric structure, starting with radio emission, pursuing with the geometry at the base of the current sheet and its related $\gamma$-ray radiation and then the X-ray emission properties. Curvature along the separatrix and along the central magnetic field line plays an important role for photon energetic and is also computed. Finally, as a direct application to existing observations, we computed multi-wavelength light-curves and extract the radio time lag variation with respect to the $\gamma$-ray peak separation for immediate comparison with 3PC.

More specifically, for the radio pulses we assume curvature emission along the open magnetic field lines delimited by the polar cap rim. Emission occurs at an altitude of $R/\rlight=0.2$. For X-rays, curvature emission happens only along the separatrix, starting at an altitude above the radio emission site and stopping before reaching the light-cylinder. These inner and outer boundaries could be variable but are prescribed by the user. Finally, for $\gamma$-rays, synchrotron emission is expecting within the current sheet of the striped wind, in the radial direction.

\subsection{Radio pulse profile and width}

\begin{figure}[h]
	\centering
	\begin{tabular}{c}
		\includegraphics[width=0.95\linewidth]{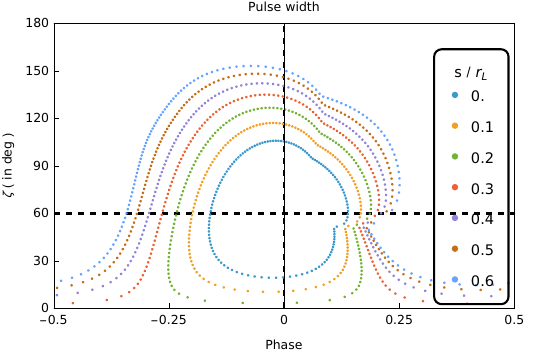} \\
		\includegraphics[width=0.95\linewidth]{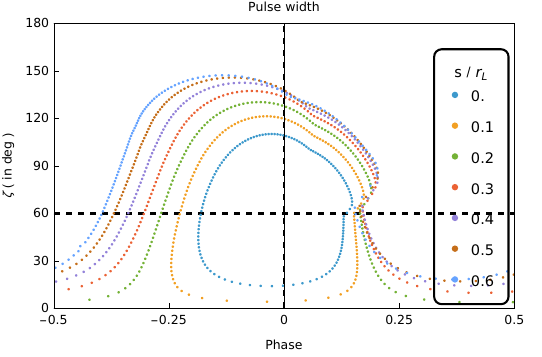} \\
		\includegraphics[width=0.95\linewidth]{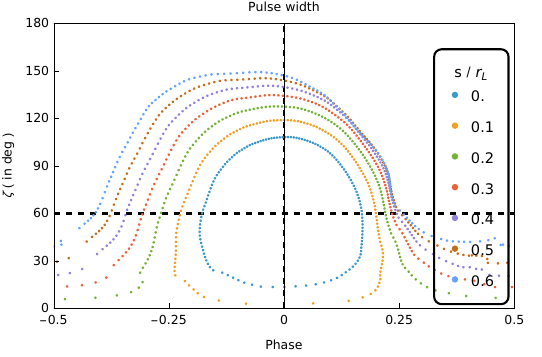}
	\end{tabular}
	\caption{Pulse with for the radio pulse profile for $\chi=60\degr$ and depending on the regime, from top to bottom: vacuum, resistive and force-free. The emission height along the field lines, starting from the surface is given by $s$ and normalised with respect to the light-cylinder radius, $s/\rlight$.}
	\label{fig:largeur_pulse_a60}
\end{figure}

Pulsars are mostly detected at low frequencies, in the radio waveband. The radio pulse profile shows a complex structure, usually very frequency-dependent. The pulse width~$W$ is related to the pulsar geometry~$\rchi$ and observer viewing angle $\zeta$ as a function of the radio beam half-opening angle~$\rho$ such that
\citep{gil_geometry_1984}
\begin{equation}\label{eq:pulse_width}
	\cos \rho = \cos \rchi \cos \zeta + \sin \rchi \sin \zeta \cos \, (W/2) \ .
\end{equation}
This expression relies on the static dipole approximation for the radio beam geometry, seen as a perfect cone. In a more realistic context, magnetic sweep-back and magnetospheric currents should be taken into account, like in the sweep-back shown in Fig.~\ref{fig:ligneschampbxyr0}. 
Fig.~\ref{fig:largeur_pulse_a60} shows sky maps of regions illuminated by the radio beam depending on the emission height~$s$ above the stellar surface and depending on the magnetospheric regime, from top to bottom: vacuum, resistive and force-free. The magnetic dipole moment inclination is fixed to $\rchi=60\degr$ and the emission height normalised to the light-cylinder radius as given in the legend by the ratio~$s/\rlight$.

In the vacuum case, we retrieve a cusp shape reminiscent of the polar cap rim. This spike disappears in the FFE regime. Note that due to the relatively large $R/\rlight$ ratio, the pulse width remains rather large, with a duty cycle of 20\%-30\%. Decreasing this ratio~$R/\rlight$ would diminish the pulse with accordingly, following expression \eqref{eq:polar_cap_radius}. An important point to notice is that large pulse widths do not necessarily imply an almost aligned rotator. Indeed, for low obliquities such as $\rchi=30\degr-45\degr$ and emission heights of $s/\rlight\approx0.5$ we still observe case with a radio pulse spanning 90\% of the period, depending on the viewing angle~$\zeta$. Consequently, millisecond pulsars detected with high radio duty cycles are compatible with mild obliquities and emission height values reaching a significant fraction of the light-cylinder radius.

\begin{figure}[h]
	\centering
	\includegraphics[width=0.95\linewidth]{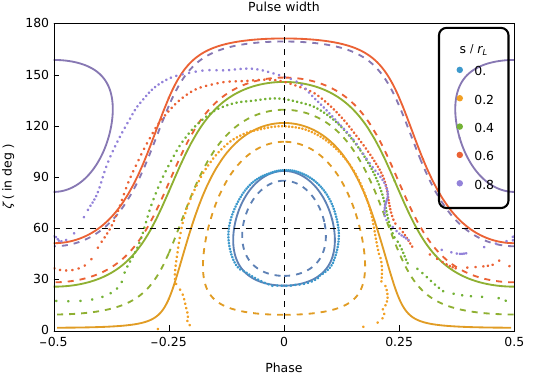}
	\caption{Sky map for the radio pulse profile for $\rchi=60\degr$ in the case of a static dipole, in dashed lines without the correcting factor~$1.2$, in solid lines with a factor~$1.2$ corrections, and a force-free dipole in dotted lines.}
	\label{fig:polar_cap_ffe}
\end{figure}

As a comparison, we show in Fig.~\ref{fig:polar_cap_ffe} the emission pattern expected from the static dipole in solid lines, in connection with the pulsar width given in Eq.~\eqref{eq:pulse_width}. Neither aberration nor retardation effects are included because we simply follow the dipole field lines, computing the local tangent~$\vec{t}$ to these lines, assuming that photons are emitted at a position marked by the spherical angles $(\theta_{\rm e}, \varphi_{\rm e})$. This leads to the following components for the tangent vector~$\vec{t}$ (see \cite{petri_multi-wavelength_2024-3} for its derivation)
\begin{subequations}
	\label{eq:tangente}
\begin{align}
	\delta_{\rm d} \, t_x & = 3 \cos \theta_{\rm e} \sin \theta_{\rm e} \cos \varphi_{\rm e} \\
	\delta_{\rm d} \, t_y & = 3 \cos \theta_{\rm e} \sin \theta_{\rm e} \sin \varphi_{\rm e} \cos \rchi + ( 3 \cos^2 \theta_{\rm e} - 1 ) \sin \rchi \\
	\delta_{\rm d} \, t_z & = ( 3 \cos^2 \theta_{\rm e} - 1 ) \cos \rchi - 3 \cos \theta_{\rm e} \sin \theta_{\rm e} \sin \varphi_{\rm e} \sin \rchi \\  
	\delta_{\rm d} & = \sqrt{3 \cos^2 \theta_{\rm e} + 1 } \ .
\end{align}
\end{subequations}
The viewing angle is therefore $\zeta = \arccos (t_z/t)$ and the azimuth $\varphi = \arctan (t_x, t_y)$. Qualitatively, the shape resembles the force-free one, but in the latter case aberration, retardation and magnetic field sweep back break the east-west symmetry. We emphasize that the above expression~\eqref{eq:tangente} also applies to the X-ray pulse profile because both radio and X-rays rely on the separatrix surface which has a simple analytical expression for the static dipole. 
Fig.~\ref{fig:polar_cap_ffe} shows also a comparison between the pulse profiles expected from the force-free magnetosphere, in dotted lines, and from the static dipole, dashed lines. For the force-free simulations, we set $R/\rlight=0.1$. At low altitude $s=0$ corresponding to the surface of the star, both profile look similar except for a larger size in the force-free, by a factor around $1.2$, see Eq.~\eqref{eq:polar_cap_radius}. For increasing emission heights, the force-free pulse is shifted to early phases due to magnetic sweep-back but the overall width remains within a factor $1.2$ of the vacuum case. Correcting for the phase shift, both regimes look qualitatively very similar at any height. However, if the factor $1.2$ is included in the vacuum polar cap shape, discrepancies arise at a height of $s/\rlight\gtrsim0.6$.

\subsection{Current sheet and $\gamma$-rays sky map}

\begin{figure}[h]
	\centering
	\includegraphics[width=0.95\linewidth]{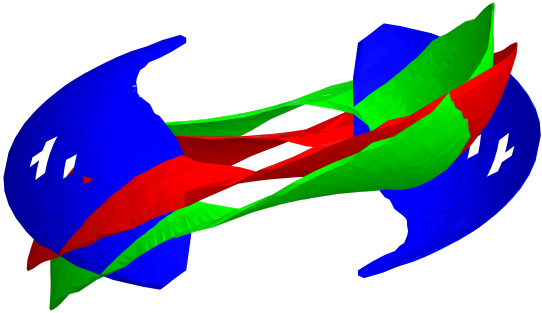}
	\caption{Three-dimensional view of the current sheet geometry for the vacuum (blue), a resistive (green) and the FFE (red) magnetosphere.}
	\label{fig:couche_3d}
\end{figure}

In the striped wind model \citep{bogovalov_physics_1999, kirk_pulsed_2002}, $\gamma$-rays are essentially produced within the current sheet outside the light-cylinder \citep{petri_unified_2011, mochol_pulsar_2017}. This surface determines the properties of the light curves: pulse shape and morphology, phase lag with respect to the radio pulse, among others. In order to understand the impact of the resistivity onto the current sheet geometry, we show some examples of this surface by plotting a three-dimensional view of this current sheet as shown in Fig.~\ref{fig:couche_3d} for $\rchi=45\degr$ in the vacuum (blue), resistive (green) and FFE (red) magnetosphere. 

\begin{figure*}[h!]
	\centering
	\includegraphics[width=0.95\linewidth]{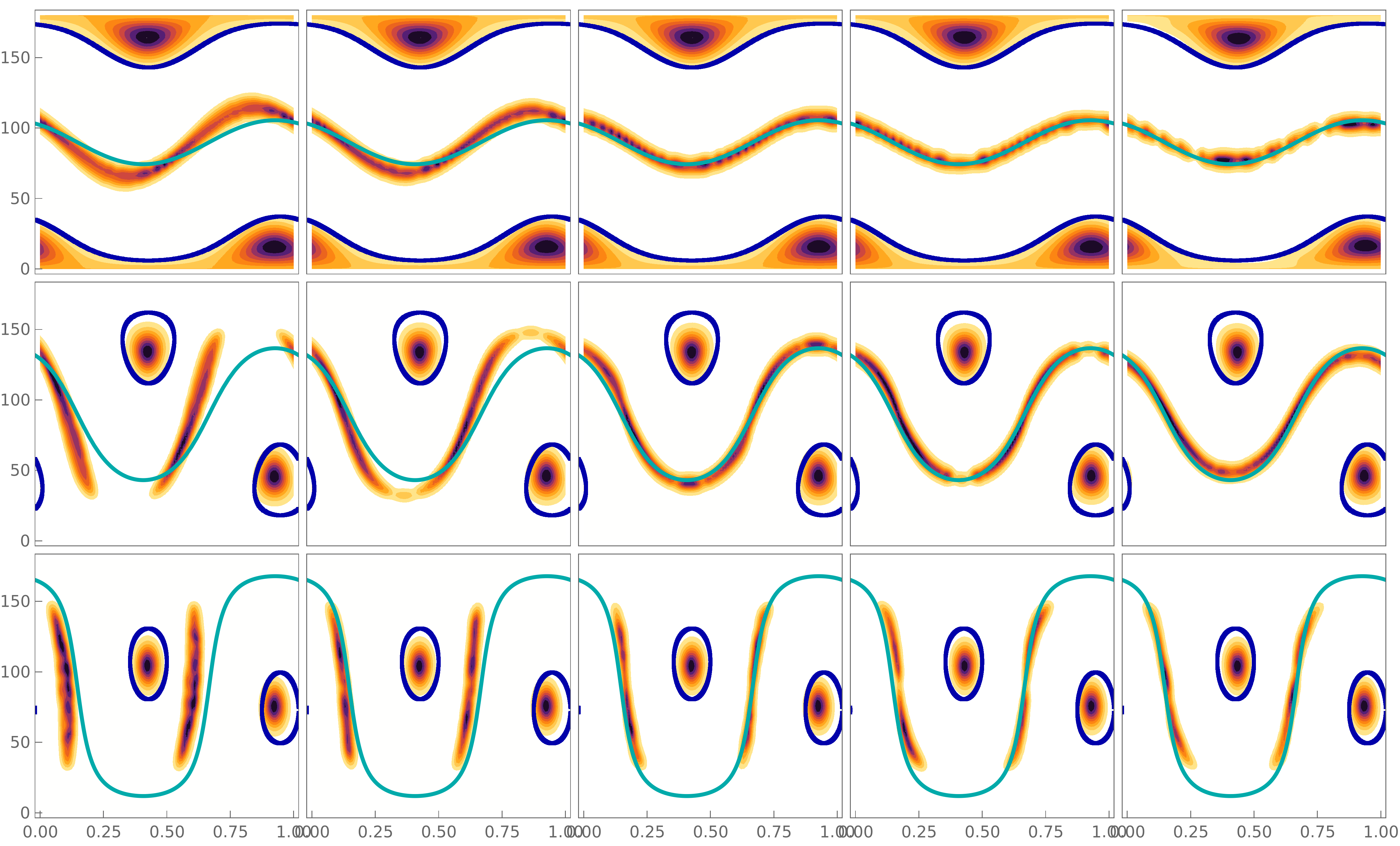}
	\caption{$\gamma$-ray sky maps for several inclinations $\rchi = \{15\degr, 45\degr, 75\degr\}$, from top to bottom. For completeness, the radio pulse profile is also shown. The conductivity increases from left to right, starting with the vacuum case and ending with the FFE regime. The blue solid lines show the sky maps deduced from a static dipole for the radio profile and the cyan solid lines show the sky maps deduced from the split monopole solution Eq.~\eqref{eq:split_monopole}.}
	\label{fig:carte_gamma}
\end{figure*}

Fig.~\ref{fig:carte_gamma} shows a complete set of $\gamma$-ray sky maps overlapped with the radio pulse profile for several inclinations $\rchi = \{15\degr, 45\degr, 75\degr\}$. In the force-free regime, the peak emission follows the split monopole pattern given by
\begin{equation}
	\label{eq:split_monopole}
	\cos \zeta = - \frac{\cos (\varphi-\varphi_0) \, \tan \rchi}{\sqrt{1 + \cos^2 (\varphi-\varphi_0) \, \tan^2 \rchi}} \ .
\end{equation}
See appendix~\ref{app:A} for the derivation of this relation.
The angle $\varphi_0$ allows for a possible phase shift in the light-curve. In the present work, it is set to ${\varphi_0 \approx -0.09}$. This curve is plotted in green lines in Fig.~\ref{fig:carte_gamma}. We see indeed a good agreement between this relation and the FFE simulations. Moving to the vacuum regime, the $\gamma$-ray peak shifts to early phases and the range of significant emission in the $\zeta$ direction also increases. This is related to the geometry of the current sheet that opens up wider in vacuum compared to FFE, see Fig.~\ref{fig:couche_3d}. Emission disappears at low and high latitude, $\zeta \lesssim 30\degr$ and $\zeta \gtrsim 150\degr$ because in our model $\gamma$-ray photons are emitted only outside the light-cylinder. Its is a purely geometric effect mostly insensitive to the obliquity~$\rchi$.

\subsection{X-rays sky map}

\begin{figure*}[h!]
	\centering
	\includegraphics[width=0.95\linewidth]{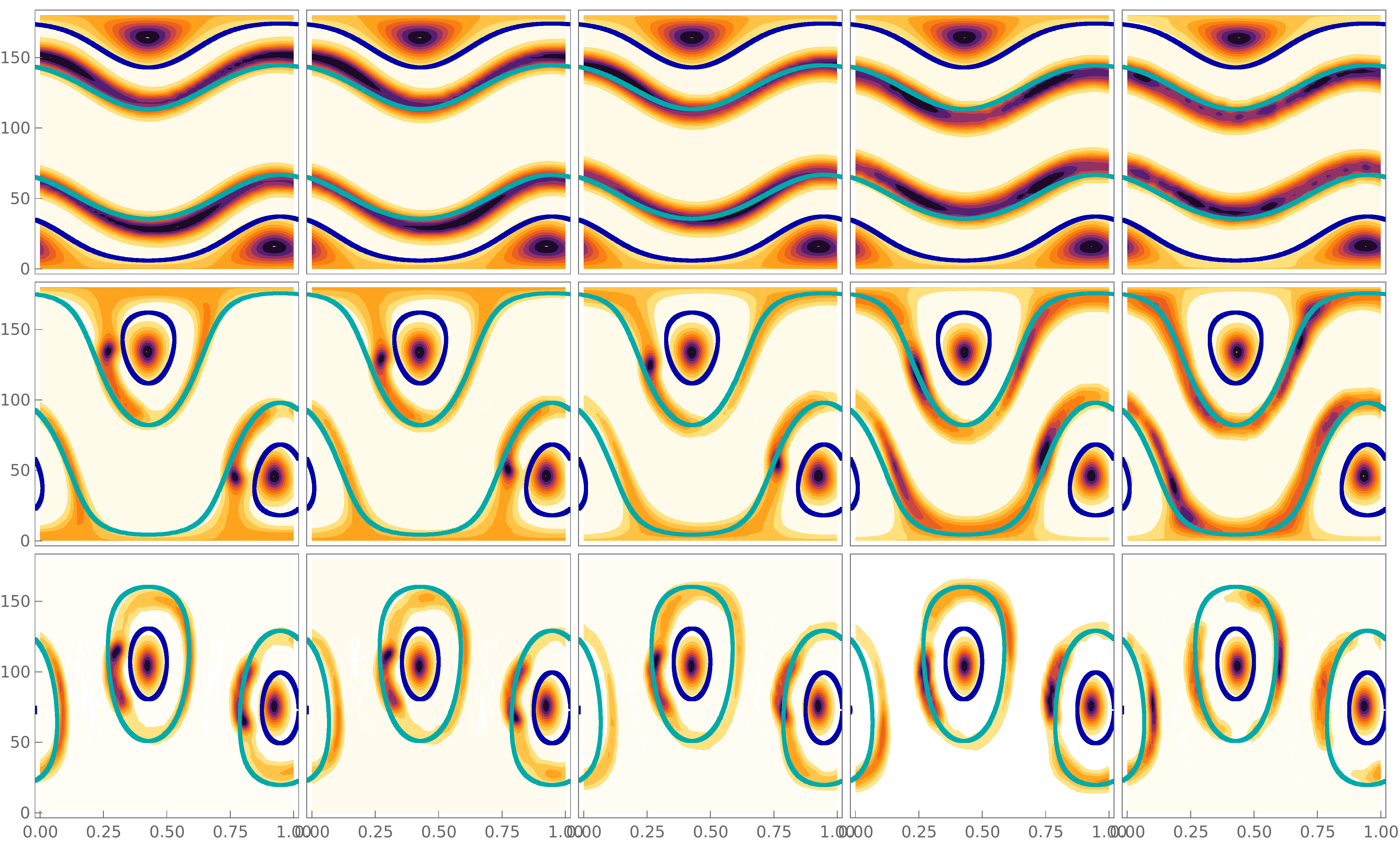}
	\caption{Same as Fig.~\ref{fig:carte_gamma} but for the X-ray sky maps. The blue solid lines show the sky maps deduced from a static dipole for the radio profile and the cyan lines for the X-ray profile.}
	\label{fig:carte_X}
\end{figure*}

The same exercise is repeated for the non-thermal X-ray produced along the separatrix. Results are shown in Fig.~\ref{fig:carte_X} and overlapped with the radio pulse profile. If the X-ray pulse profiles are phase-aligned with the radio pulse, decreasing the conductivity shifts the X-ray pulses to later phases. This behaviour is opposite to the $\gamma$-ray light-curve expectations. The variations in pulse shape are drastic in X-rays. Indeed, whereas in the FFE case, the peak in X-ray lags the radio pulse, in low conductivity and vacuum regimes, this X-ray peak leads the radio profile. See for instance Fig.~\ref{fig:courbe_gamma_a75_z60} for a typical example of peak changing in X-rays.

As a consequence, the most visible impact of the magnetosphere resistivity resides in shaping the non-thermal X-ray emission, and not the radio and/or $\gamma$-ray energy band.
The X-ray sky maps suppose that photons are emanating from the separatrix. In this particular example, emission starts at $R/\rlight=0.2$ and stops at $R/\rlight=0.3$. Other emission heights are possible and would lead to different X-ray sky maps, see for instance \cite{petri_multi-wavelength_2024-3} for other examples.

\subsection{Curvature along the magnetic axis and the separatrix}

Photons within the magnetosphere are produced via synchrotron and curvature radiation. The typical synchrotron energy is imposed by the local magnetic field strength whereas the curvature energy depends linearly on the local curvature of the magnetic field lines. In this section we are interested in the latter radiation mechanism. Two kinds of field lines are particularly important: magnetic field lines sustaining the separatrix and the one along the magnetic axis.

\begin{figure}[h]
	\centering
	\includegraphics[width=0.95\linewidth]{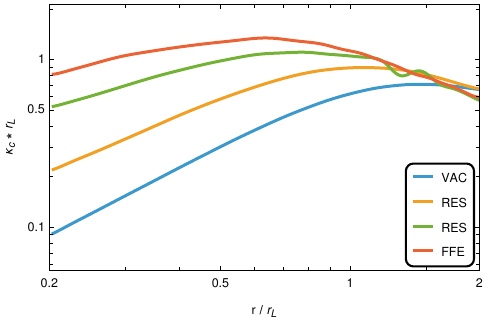}
	\caption{Curvature along the central magnetic field line for vacuum, resistive and FFE with obliquity $\rchi=60\degr$.}
	\label{fig:courbure_centrale_vide_res_ffe_a60}
\end{figure}

\begin{figure}[h]
	\centering
	\includegraphics[width=0.95\linewidth]{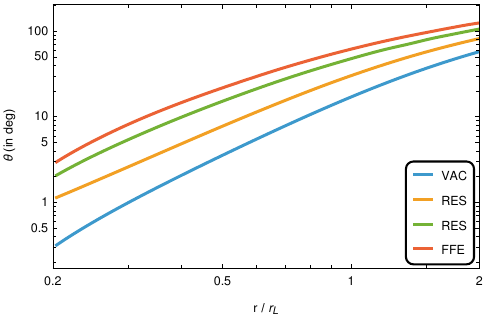}
	\caption{Angle between the magnetic axis and the central magnetic field line for vacuum, resistive and FFE with obliquity $\rchi=60\degr$.}
	\label{fig:angle_tx_centrale_vide_res_ffe_a60}
\end{figure}

Fig.~\ref{fig:courbure_centrale_vide_res_ffe_a60} shows the curvature~$\kappa_{\rm c}$ in units of $1/\rlight$ for several plasma regimes when sliding along the magnetic axis. In all cases, the curvature increase from the surface to an altitude of $R/\rlight\approx 0.5-1$ and then decreases. The FFE regime offers the highest curvature, which is expected since the electric current bends field lines stronger compared to vacuum or resistive cases. If radio photons are produced by curvature radiation along field lines around the magnetic axis, we should expect to observe low frequency at low altitude and high frequency at high altitude, the exact opposite to the radius-to-frequency mapping (RFM) findings \citep{komesaroff_possible_1970}. This frequency evolution with altitude, thus with curvature, relies on the fact that all emitting particles possess the same Lorentz factor~$\gamma$, i.e. a mono-energetic particle distribution function, irrespective of their height above the polar cap, along the magnetic axis, because the typical curvature radiation frequency is given by $\omega_{\rm curv} \approx (3/2) \, \gamma^3 \, c \, \kappa_{\rm c}$. From this expression, we immediately notice than an increase in curvature $\kappa_{\rm c}$ leads to a  similar increase in frequency $\omega_{\rm curv}$.
The associated angle between the magnetic axis and the tangent to the central magnetic field line is shown in Fig.~\ref{fig:angle_tx_centrale_vide_res_ffe_a60}. It is largest for the FFE case and diminishes for the resistive and vacuum cases. In the vacuum case, the central field line remains almost straight up to the light-cylinder, where it shows a deviation less than $20\degr$. This needs to be contrasted with the FFE line deviating more than $60\degr$.

\begin{figure}[h]
	\centering
	\includegraphics[width=0.95\linewidth]{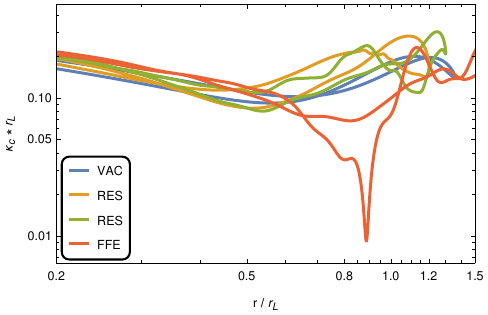}
	\caption{Curvature along the separatrix for vacuum, resistive and FFE with obliquity $\rchi=60\degr$.}
	\label{fig:courbure_separatrice_vide_res_ffe_a60}
\end{figure}

\begin{figure}[h]
	\centering
	\includegraphics[width=0.95\linewidth]{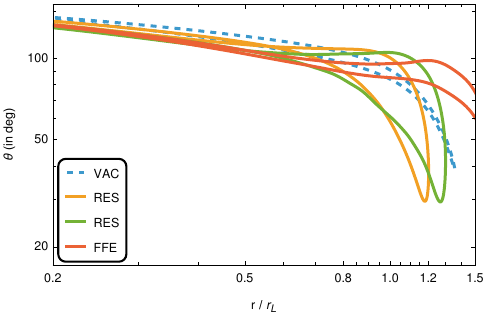}
	\caption{Angle between the magnetic axis and the tangent to the local magnetic field line along the separatrix for vacuum, resistive and FFE with obliquity $\rchi=60\degr$.}
	\label{fig:angle_tx_separatrice_vide_ffe_a60}
\end{figure}

In the separatrix region, the situation reverses. Fig.~\ref{fig:courbure_separatrice_vide_res_ffe_a60} indeed shows the curvature~$\kappa_{\rm c}$ in units of $1/\rlight$ for several plasma regimes when sliding along field lines, starting at the surface of the star, close to the north pole, reaching a maximal distance when grazing the light-cylinder, and then turning back to the stellar surface, at the south pole. In this back and forth travel, two lines appear for the same radial coordinate~$r$, one for the leading direction (forward motion) and the other for the receding direction (backward direction). In all cases, the curvature decrease from the surface to an altitude of $R/\rlight\approx 0.5$ and then increases again up to the point where the magnetic field line returns to the surface. The configuration is not symmetric with respect to the turning point as seen in the plot. Moreover, the FFE regime offers the highest curvature, which is expected since the electric current bends field lines stronger compared to vacuum or resistive cases. If radio photons are produced by curvature radiation along the separatrix, we would expect to observe low frequency at high altitude and high frequency at low altitude, this time agreeing with the radius-to-frequency mapping predictions.
The associated angle between the magnetic axis and the tangent to the local magnetic field line along the separatrix is shown in Fig.~\ref{fig:angle_tx_separatrice_vide_ffe_a60}. It is largest for the vacuum case and diminishes for the resistive and FFE cases.

\subsection{Multi-wavelength light-curves}

\begin{figure}[h]
	\centering
\begin{tabular}{c}
	\includegraphics[width=0.95\linewidth]{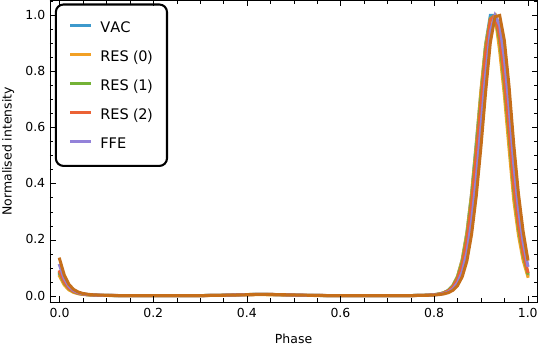} \\
	\includegraphics[width=0.95\linewidth]{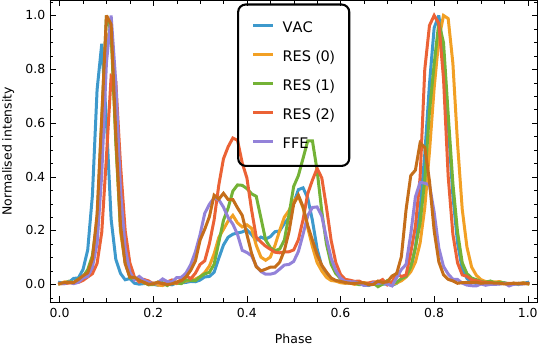} \\
	\includegraphics[width=0.95\linewidth]{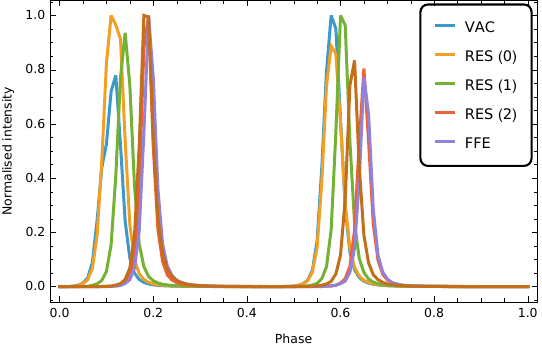}
\end{tabular}
	\caption{Radio, X-ray and $\gamma$-ray light-curves, from top to bottom, for $\rchi=75\degr$ and $\zeta=60\degr$ in the different magnetosphere models.}
	\label{fig:courbe_gamma_a75_z60}
\end{figure}

Radio emission is generated well inside the light-cylinder, we therefore do not expect to observe a dramatic impact of the resistivity onto the radio pulse profiles. However, for the X-ray and $\gamma$-ray emissions, as these are generated around the light-cylinder where the magnetic field geometry is strongly impacted by the resistivity, we will observe a significant variation in the light curves. Fig.~\ref{fig:courbe_gamma_a75_z60} shows an example of multi-wavelength pulse profile variation due to the plasma feedback. The light-curves are normalised to a maximum intensity of~$1$. The radio profile is not impacted by the resistivity, as expected. Neither the shape nor the phase is changed appreciably. However, in X-rays, the pulse profiles drastically depend on the plasma regime. The number of peaks and their relative amplitudes vary with the conductivity parameter~$\tilde{\sigma}$. Finally in $\gamma$-ray, the number of peaks is preserved, they are shifted and their relative amplitude also varies. The most important conclusion to draw from this section is that the radio pulses are not affected by the plasma current because photons are generated deep inside the magnetosphere where the static dipole geometry remains a good approximation.
The same remark applies to the $\gamma$-ray emission except for a possible shift in phase of the profile. These two extreme wavebands offer a robust way to constrain the pulsar geometry, irrespective of the assumption about the particle content of the magnetosphere. Only X-ray and to a lesser extent $\gamma$-rays could give some hints to the conductivity value~$\tilde{\sigma}$.

\begin{figure}[h]
	\centering
	\includegraphics[width=0.95\linewidth]{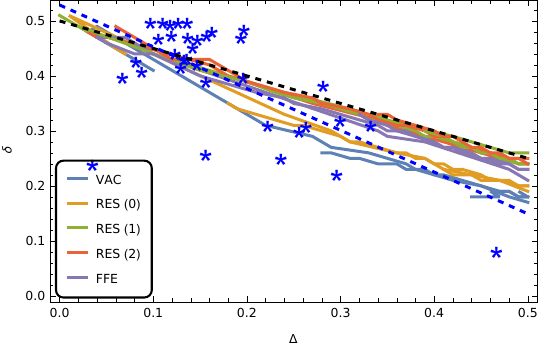}
	\caption{Radio time lag~$\delta$ versus $\gamma$-ray peak separation~$\Delta$. Blue stars represent data for young pulsars from the 3PC. The black dashed line represents the theoretical prediction Eq.~\eqref{eq:deltavsDELTA_theorie} and the blue dashed line a linear fit to the data, Eq.~\eqref{eq:deltavsDELTA_fit}.}
	\label{fig:delta_vs_delta}
\end{figure}

As a final summary of the plethora of multi-wavelength pulse profiles that we cannot expose here, we show the radio time lag~$\delta$, i.e. the temporal delay between the first (or unique) $\gamma$-ray peak and the main radio pulse profile, versus the $\gamma$-ray peak separation~$\Delta$ for young pulsars, meaning with periods $P>30$~ms, in blue stars in Fig.~\ref{fig:delta_vs_delta}. Data are taken from the 3PC catalogue \citep{smith_third_2023}. We also plot the expectation from our model in the different plasma regimes, from vacuum, through resistive up to FFE magnetospheres in colour solid lines in the same Fig.~\ref{fig:delta_vs_delta} (see also \cite{contopoulos_pulsar_2010} for similar results in FFE). The $\delta-\Delta$ relation is rather insensitive to the plasma regime, excepted for an almost empty magnetosphere where the radio time lag decreases by 5\% to 10\%. Interestingly, this relation remains independent of the angles $\rchi$ and $\zeta$. In other words, the geometry of the pulsar needs not to be known. For high plasma conductivity and in FFE, we retrieve to good accuracy the relation
\begin{equation}
	\label{eq:deltavsDELTA_theorie}
	\delta \approx \frac{1-\Delta}{2} = 0.5 - 0.5 \, \Delta
\end{equation}
as derived in \cite{petri_unified_2011}. This theoretical expression is shown in black dashed line in Fig.~\ref{fig:delta_vs_delta}. Unfortunately, this time lag is therefore not a discriminating parameter to constrain the plasma conductivity. However, in all cases, it seems that our model overestimates the time lag for large peak separations satisfying $\Delta\gtrsim0.3$, although the number of observations in this range is extremely scarce. A straightforward linear fit to the observations indeed gives
\begin{equation}
	\label{eq:deltavsDELTA_fit}
	\delta \approx 0.53 - 0.76 \, \Delta
\end{equation}
which is shown in blue dashed line in Fig.~\ref{fig:delta_vs_delta}. As shown in \cite{petri_multi-wavelength_2024-3}, a non purely radial outflow at the light-cylinder can decrease the time lag to reconcile with observations. PIC simulations and theoretical work by  \cite{contopoulos_hybrid_2020} indeed show that a significant velocity component exists along the azimuthal direction outside the light-cylinder, and being tangent to the light-cylinder surface. Such freedom would certainly better fit the data but at the expense of adding a free parameter to fix the angle between the radial direction and the velocity vector. 

\subsection{Beaming factor}

\begin{figure*}[h!]
	\centering
	\begin{tabular}{ccc}
	\includegraphics[width=0.33\linewidth]{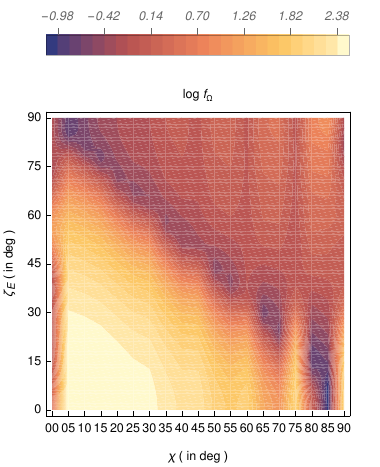} &
	\includegraphics[width=0.33\linewidth]{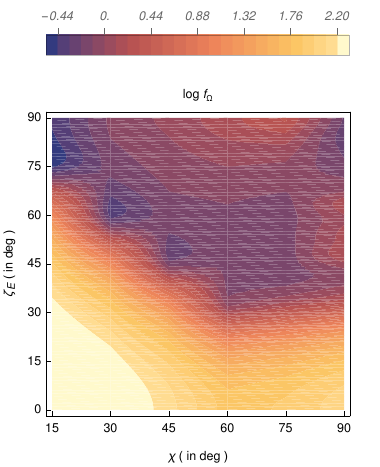} &
	\includegraphics[width=0.33\linewidth]{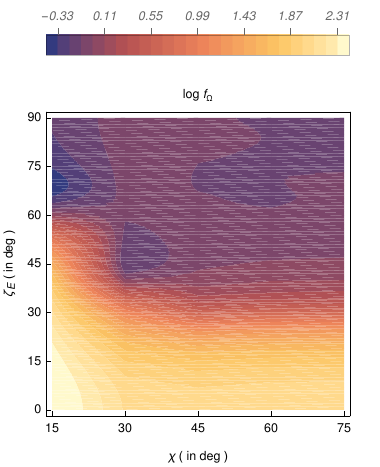}
	\end{tabular}
	\caption{Logarithm of the beaming factor $f_\Omega$, for the force-free magnetosphere on the left panel, for the resistive case with $\log \tilde{\sigma}=2$ on the middle panel, and for vacuum on the right panel.}
	\label{fig:FacteurCorrection}
\end{figure*}

A last interesting quantity to compare is the beaming factor $f_\Omega$, useful to compute the total $\gamma$-ray luminosity $L_\gamma$ in the striped wind model such that
\begin{equation}
	\label{eq:LuminositeGamma}
	L_\gamma = 4 \, \pi \, f_\Omega \, F_{\rm obs} \, D^2 \ .
\end{equation}
Here $D$ is the distance of the pulsar to the observer and $F_{\rm obs}$ the observed flux. The factor $f_\Omega$ is defined from the phase-resolved $\gamma$-ray flux $F_\gamma(\rchi,\zeta,\varphi)$ by \cite{watters_atlas_2009} as
\begin{equation}
	\label{eq:FacteurFocal}
	f_{\Omega}(\rchi,\zeta_E) = \frac{\int_0^{\pi} \int_0^{2\,\pi} F_\gamma(\rchi,\zeta,\varphi) \, \sin\zeta \, d\zeta \, d\varphi}
	{2 \, \int_0^{2\,\pi} F_\gamma(\rchi,\zeta_E,\varphi) \, d\varphi}
\end{equation}
For the striped wind model, this correction factor is shown in fig.~\ref{fig:FacteurCorrection} with the full dependence on obliquity $\chi$ and inclination of Earth line of sight $\zeta_E$.

\subsection{Observational impact of the resistivity}

As we have seen, the resistivity has a strong impact on the shape of the separatrix, Fig.~\ref{fig:separatrice}. The result is a strong variation in the X-ray pulse profile as shown in the middle panel of Fig.~\ref{fig:courbe_gamma_a75_z60}. Moreover, in a previous work, \cite{petri_localizing_2024} showed that this separatrix could be responsible for the non-thermal X-ray emission of pulsars. If this is confirmed in other pulsars, observations of non-thermal X-rays with its phase-resolved polarimetry would drastically constrain the magnetic field geometry and height of these emission sites. Unfortunately, pulsars bright enough in X-rays are rare and the current X-ray telescopes like IXPE (Imaging X-ray Polarimetry Explorer) are not sensitive enough to measure this polarisation.

Another approach consist of simulating pulsar populations of radio-loud and radio quiet $\gamma$-ray pulsars and to check which resistive model better reproduces the observations as done for instance a decade ago by \cite{pierbattista_constraining_2012,pierbattista_light-curve_2015,pierbattista_young_2016}. However, performing pulsar populations syntheses requires numerous input parameters for the birth, evolution and detection of pulsars, as described in \cite{sautron_galactic_2024}. Such study would deserve a fully detailed analysis that largely goes beyond the scope of this work.

\section{Conclusions\label{sec:Conclusion}}

The precise plasma content of pulsar magnetospheres is still highly debated. Does it contain only leptonic pairs or also protons and maybe ions? And where should these particles be injected? This plasma feeds back to the magnetic field structure and should lead to unambiguous observational signatures. However, we demonstrated that the radio signal is rather insensitive to the plasma resistivity. Only the X-ray and $\gamma$-ray pulse profiles are impacted by the plasma composition. The polar cap geometry remains similar in shape and size irrespective of the conductivity~$\sigma$. Nevertheless, the spin down luminosity as well as the magnetic field sweep-back effect in the vicinity of the light-cylinder are affected, reflecting into the $\gamma$-ray pulse profiles and phase alignment with the radio pulse. However, we only found a weak dependence in the relation between the $\gamma$-ray peak separation~$\Delta$ and the radio time lag~$\delta$.

The above approach suffers from some limitations, notably the assumption of a spatially constant and uniform conductivity~$\sigma$ within the magnetosphere. Regions around the neutron star could certainly be split into places where the ideal plasma regime without dissipation holds and other places where dissipation occurs at a variable rate, not necessarily well described by a conductivity~$\sigma$. PIC simulations are able to handle simultaneously and self-consistently particle acceleration and radiation, but unfortunately they can not yet handle strong magnetic field strengths of the order of $10^8$~T. However, existing PIC simulations, event without a realistic scaling, could give some hints on the resistivity, informing us on the actual value of $\sigma$ as a function of location within the magnetosphere. Force-free Inside Dissipative Outside (the so-called FIDO model) \citep{kalapotharakos_gamma-ray_2014} or other approximations like resistive magnetospheres are good alternatives to grasp the full dynamics of relativistic and strongly magnetised magnetospheres. To go further into the discrimination between resistive magnetospheres, we need to consider the dynamics of particle acceleration to ultra-relativistic speeds, their radiation and energetics, because curvature is strongly affected by the magnetospheric current, producing different spectra and light-curves.

\begin{acknowledgements}
I am grateful to the referee for helpful comments and suggestions. This work has been supported by the grant number ANR-20-CE31-0010.
\end{acknowledgements}

%\bibliographystyle{aa}
%\bibliography{/home/petri/zotero/Ma_bibliotheque}

\appendix

\section{Split monopole approximation\label{app:A}}

\begin{figure}[h!]
	\centering
	\includegraphics[width=0.9\linewidth]{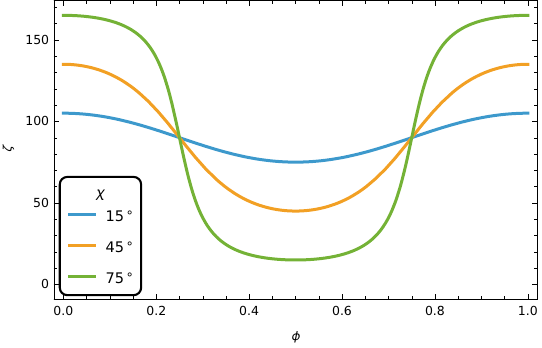}
	\caption{Loci of the $\gamma$-ray emission for the split monopole model for obliquities ${\rchi = \{15\degr, 45\degr, 75\degr\}}$ and $\varphi_0 = 0$.}
	\label{fig:split_monopole}
\end{figure}

Let us assume that the magnetic axis is given by 
\begin{equation}\label{key}
	\vec{\mu} = \sin \rchi \, \ex + \cos \rchi \, \ez
\end{equation}
in a Cartesian orthonormal basis denoted by $(\ex, \ey, \ez)$. The current sheet is perpendicular to this vector, therefore any vector $\vec{n}$ pointing to the current sheet must satisfy $\vec{\mu} \cdot \vec{n} = \vec{0}$. Its plane is defined by
\begin{equation}
	x \sin \, \rchi + z \, \cos \, \rchi = 0 \ .
\end{equation}
To find the curve intersecting the light cylinder, we furthermore impose $x^2 + y^2 = \rlight^2$ thus we write $x = \rlight \cos \varphi$ and $y = \rlight \sin \varphi$. The vector position $\vec{r}$ of any point on the ellipse defined by the intersection of the current sheet with the light cylinder satisfies 
\begin{equation}
	\vec{r} = \cos \varphi \, \ex + \sin \varphi \, \ey - \cos \varphi \tan \, \rchi \, \ez \ .
\end{equation}
This vector is not normalised because $\|\vec{r}\|^2 = 1 + \cos^2 \varphi \tan^2 \, \rchi$. If emission is radial, then the sky map is defined by a curve in the $(\varphi,\zeta)$ plane such that
\begin{subequations}
	\begin{align}
	\varphi & = \arctan(r_x, r_y) \\
	\zeta & = \arccos \frac{r_z}{r} = \arccos \left( \frac{- \cos \varphi \tan \, \rchi}{\sqrt{1 + \cos^2 \varphi \tan^2 \, \rchi}} \right) \ .
	\end{align}
\end{subequations}
In order to allow for a possible phase shift $\varphi_0$, we use the transformation $\varphi \rightarrow \varphi - \varphi_0$. Fig.~\ref{fig:split_monopole} shows some examples of loci of this curve $\zeta(\varphi)$ in a sky map diagram for obliquities $\rchi = \{15\degr, 45\degr, 75\degr\}$ and $\varphi_0 = 0$.

\section{Current sheet dynamics}
\begin{figure}[h!]
	\centering
	\includegraphics[width=0.9\linewidth]{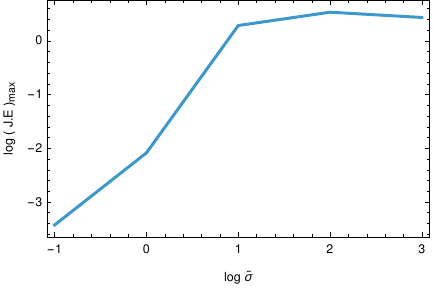}
	\caption{Maximum dissipation power in the current sheet outside the light-cylinder in arbitrary units. It saturates for high conductivity and decreases for low conductivity, vanishing for the vacuum field.}
	\label{fig:dissipation_max}
\end{figure}

\begin{figure*}[h]
	\centering
	\includegraphics[width=0.33\linewidth]{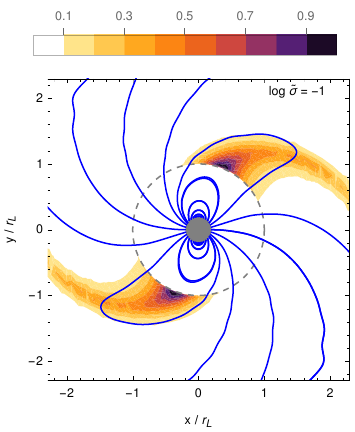}
	\includegraphics[width=0.33\linewidth]{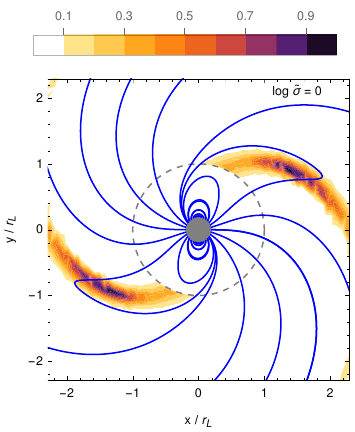}
	\includegraphics[width=0.33\linewidth]{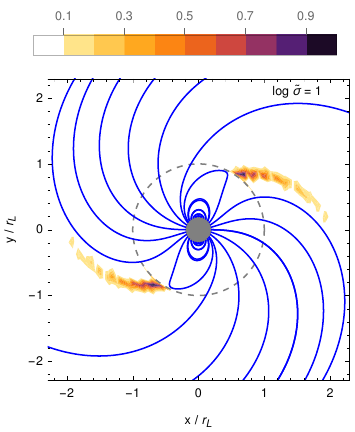}
	\caption{Isocontours of the power $\vec{j} \cdot \vec{E}$ dissipated in the current sheet outside the light-cylinder for a perpendicular rotator with conductivity ${\log \tilde{\sigma} = \{-1,0,1\}}$ from left to right. The maximum dissipation is normalised to unity in all cases for better comparison. The gray disk depicts the neutron star and the gray dashed circle the light-cylinder.}
	\label{fig:jse_equateur_a90_j5}
\end{figure*}

In our simulations, the current sheet is not resolved in the force-free regime as it is theoretically of zero thickness. Even in the radiative regime, it remains at a level below the resolution of our numerical grid. However, for mildly to highly resistive solutions, the current sheet is resolved within several grid points. The maximum dissipation power computed through $\vec{j} \cdot \vec{E}$ in the current sheet outside the light-cylinder is shown in Fig.~\ref{fig:dissipation_max} for all values of the conductivity $\tilde{\sigma}$. It obviously vanishes for the vacuum rotator and increases with increasing conductivity $\sigma$, saturating for very high conductivity $\tilde{\sigma} \gg 1$.
Radiation emanates from regions where dissipation is important. It is therefore crucial to localise where Joule heating occurs within the magnetosphere by computing the term $\vec{j} \cdot \vec{E}$ in whole space. In the force-free regime, this term vanishes identically in space by construction and in the vacuum case, it also vanishes identically because of the absence of any electric current. Only the radiative and resistive models are meaningful for studying dissipation. 
The region of significant dissipation outside the light-cylinder, given by the extremum of the power $\vec{j} \cdot \vec{E}$ is shown in Fig.~\ref{fig:jse_equateur_a90_j5} for conductivity ${\log \tilde{\sigma} = \{-1,0,1\}}$ from left to right. The maximum value of the power for each plot is normalised to unity for better comparison of the current sheet dissipation layer thickness. We notice that the thickness of the region of significant dissipation increases when the conductivity~$\tilde{\sigma}$ decreases, potentially impacting the width of the $\gamma$-ray pulse profiles but not their number and overall shape. However, from the location of these dissipation layers, we do not expect a significant change in the light curve phase positions.

\begin{figure*}[h]
	\centering
	\includegraphics[width=0.95\linewidth]{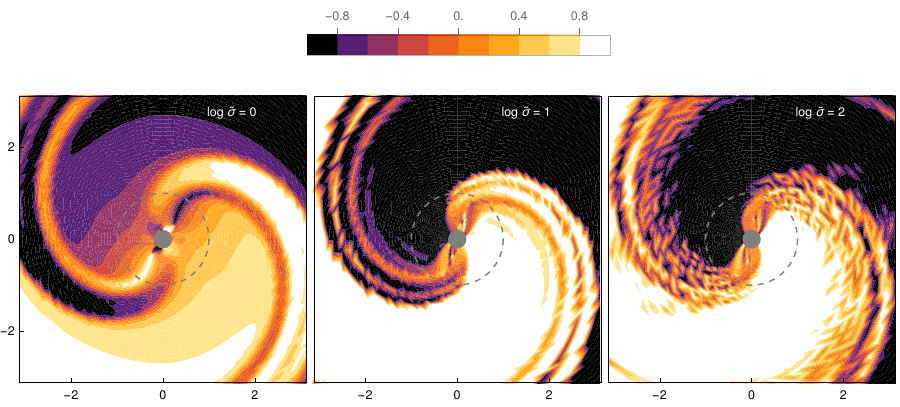}
	\caption{Cosinus of the angle between the direction of the current and the radial direction for a perpendicular rotator with resistivity ${\log \tilde{\sigma} = \{0,1,2\}}$ from left to right. The gray disk depicts the neutron star and the gray dashed circle the light-cylinder.}
	\label{fig:cos_j_equateur_a90_j5}
\end{figure*}

The current is known in the simulations with sufficient resistivity but its direction is not necessarily straightforwardly associated to the direction of the plasma flow or to the particle velocity, except in the fully charge-separated cases. Actually, \cite{li_resistive_2012} showed that the plasma fluid velocity is along the $\vec{E} \wedge \vec{B}$ drift direction, thus no motion along the magnetic field.
Fig.~\ref{fig:cos_j_equateur_a90_j5} shows that the current is almost perfectly radial outside the light-cylinder, except for significant resistivity where the current possesses a non negligible azimuthal component. So far, field lines have been used as a proxy to localise the current sheet where emission occurs but another prescription could be related to the dissipation depicted by the power $\vec{j} \cdot \vec{E}$. This leads to a complementary way to localise the current sheet as shown in Fig.~\ref{fig:jse_equateur_a90_j5}. The dissipation layer location in space is consistent with the determination of the current sheet region solely by the magnetic field geometry.

\end{document}